\documentclass[sigconf]{acmart}
\usepackage{makecell}
\usepackage{caption}
\usepackage{subcaption}
\usepackage{blindtext}

\AtBeginDocument{%
  \providecommand\BibTeX{{%
    \normalfont B\kern-0.5em{\scshape i\kern-0.25em b}\kern-0.8em\TeX}}}

\setcopyright{acmcopyright}

\copyrightyear{2023}
\acmYear{2023}
\setcopyright{rightsretained}
\acmConference[CHI EA '23]{Extended Abstracts of the 2023 CHI Conference on Human Factors in Computing Systems}{April 23--28, 2023}{Hamburg, Germany}
\acmBooktitle{Extended Abstracts of the 2023 CHI Conference on Human Factors in Computing Systems (CHI EA '23), April 23--28, 2023, Hamburg, Germany}\acmDOI{10.1145/3544549.3585772}
\acmISBN{978-1-4503-9422-2/23/04}

\begin{document}


\title{Print Your Money: Cash-Like Experiences with Digital Money}

\author{Chenhang Zhou}
\email{zhouche@student.ethz.ch}
\affiliation{%
  \institution{ETH Zurich}
  \city{Zurich}
  \country{Switzerland}
}

\author{Yu Chen}
\email{yuchen14@student.ethz.ch}
\affiliation{%
  \institution{ETH Zurich}
  \city{Zurich}
  \country{Switzerland}
}

\author{Roger Wattenhofer}
\email{wattenhofer@ethz.ch}
\affiliation{%
  \institution{ETH Zurich}
  \city{Zurich}
  \country{Switzerland}
}

\author{Ye Wang}
\email{wangye@um.edu.mo}
\affiliation{%
  \institution{University of Macau}
  \city{Macau}
  \country{China}
}

\begin{abstract}
The use of digital money has become increasingly popular, but it comes with certain drawbacks. For instance, it can be challenging to make payments during power outages or internet failures. Additionally, some groups may find it difficult to use digital money. To address these concerns, we propose a design for a central bank digital currency (CBDC) similar to physical cash but also integrates with digital payment systems. This would enable users to access digital money without needing a third party. Our design also addresses technical and security concerns by implementing a trust-level model and ensuring that the system meets users' security needs. Ultimately, our design has the potential to replace physical banknotes and coins.

\end{abstract}



\maketitle

\section{Introduction}

The way people transact has been evolving over time. Currently, we are in the midst of another pivot point where physical money is being replaced by digital forms of payment, such as apps and digital money. Countries leading digital payments, such as Sweden and China, can serve as valuable examples to learn from as the shift towards digital payments is happening globally. These countries have already seen widespread adoption of digital payments, with many stores no longer accepting cash~\cite{shen2020can, arvidsson2018nar,arvidsson2019building}. In a study by the Swedish Retail and Wholesale Council, more than half of the retail businesses expect to stop accepting cash after 2025~\cite{arvidsson2018nar,arvidsson2019building}.

Digital currency can take several forms. It could be issued by companies, such as Facebook's Diem project. More likely, our future money will be Central Bank Digital Currency (CBDC), a digital currency issued by a country's central bank~\cite{bech2017central, bjerg2017designing}. Many central banks are already pushing ahead with the development of CBDC~\cite{boar2021ready,riksbank2017riksbank,licandro2018uruguayan,mu2019}. The user experience of CBDC will probably be similar to payment apps or credit/debit cards. Payments will be contactless, and transactions between private individuals will be possible.

The majority of central banks believe that CBDC will eventually be available to the public, and more than 75\% of central banks have already been engaging in CBDC~\cite{boar2021ready}. However, the implementation of CBDCs is not without difficulties. One major concern is how to facilitate transactions during contingencies, such as when the power or internet is down. In these scenarios, it should still be possible to complete transactions, but this raises the potential for fraud if the system is not properly secured~\cite{shen2020can}. Another challenge with CBDCs is that everyone who participates in the system must have a CBDC account, which could exclude certain groups such as children, the elderly, and non-resident travelers.

These examples show that a backup is necessary. For a while, cash will be this backup solution. But running an orthogonal second payment system is not a valid long-term strategy. If cash is not used on a regular basis, even its security will be in jeopardy. People will not recognize the security features of rarely seen banknotes, and hence the trust in banknotes will vanish. Once cash is at this stage, it will not be helpful in emergencies. So central banks will stop producing cash altogether since the production and logistics costs of cash are high. We believe that cash may be disappearing surprisingly quickly, first in countries like Sweden, then in other parts of the world~\cite{arvidsson2019building}.

In principle, cash and CBDC represent two payment systems that are essentially incompatible. In this paper, we propose a cash-like alternative that can be fully compatible with CBDC and, as such, better suited for the future of digital payments. Our design process has three stages: interviews, system design, and system tests. First, we conducted 15 interviews in China, where digital payment systems are predominant, to acquire insights into how a CBDC system should satisfy their different needs. Based on these interviews, we developed a prototype CBDC system. Finally, we tested the prototype's performance on four different devices and invited six users to experience it in different simulated scenarios.

 
\section{Preferences for Digital Money and Cash}

\begin{figure*}
    \centering
    \includegraphics[width=\textwidth]{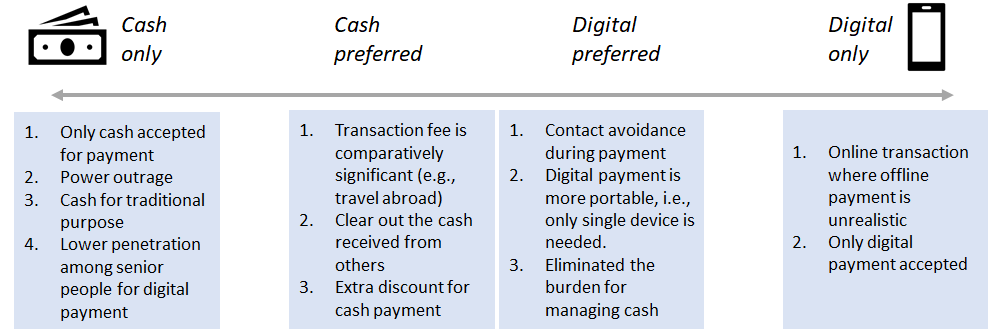}
    \caption{Potential reasons why people choose cash or digital payment from literature.}
    \label{fig:spe}
\end{figure*}

To design a digital money system that meets people's needs, we first study how people perceive current payment systems. In particular, user preference for cash or digital payments can inspire us to determine the features of our designed system. There are a series of works that have investigated people's payment preferences with empirical methods~\cite{ferreira2014building,Moneytalk2014Finance,shen2020can,kishi2019project,vines2012joy,monica2017friends, perry2018moneywork,Ghosh2015Ghana,Gupta2014IndiaTurk,John2014ManagingLowIncome,Mehra2018PrayanaIF}, especially their choices between cash and digital payments~\cite{ferreira2014building,Ferreira2019SocialConsideration,perry2018moneywork, Ferreira2017Spending}.
We review previous literature and categorize user payment preferences as four scenarios accordingly: cash-only, cash-preferred, digital-preferred, and digital-only (cf. Figure~\ref{fig:spe}). 

\subsection{Cash-only Reasons}
Intuitively, some people ``persist'' using cash as they are unbanked, either because of age, illiteracy, or perceived unsuitability~\cite{kumar2011times}, while digital payment systems linked with banking account are always mandatory besides a few exceptions~\cite{Unbanked2014financial}. Kumar et al.~\cite{kumar2011times} detailed scenarios where some Indian vendors are unbanked due to reasons such as long travel distance to the bank and perceived inconvenience of being a bank user.
On top of this, payees' technical constraint is still a determinant that forces people to choose cash without alternatives. For example, under catastrophes that lead to power or communication signal temporary outages, paralyzing digital payment infrastructure~\cite{shen2020can}, vendors and merchants are left with accepting cash, which results in cash-only scenarios for payers as well.

\subsection{Cash-preferred Reasons}
Other than cash's dominance in the cash-only cases above, reasons supporting cash-preferred scenarios where people are prone to use cash, even when digital payment is available, are presented in many works. 
Financially speaking, payment and transaction via cash rarely incur additional service fees, which digital payment likely does. Traveling abroad and cross-border transaction fees are always excessively high and will hardly make economic sense to the budget~\cite{Lehdonvirta2009TransactionCosts}. Therefore, people may prefer to use cash in such scenarios. Also, in terms of financial control, as elaborated by Joseph at al.~\cite{ratan2010finance} that cash, checks, and coupons will assist people in tracking and controlling their personal finance efficiently~\cite{Moneytalk2014Finance}.

Another remarkable attribute of cash, compared with digital payment, is that cash is tangible~\cite{o2017working} and socially symbolic~\cite{wu2017redpacket}, enticing people to choose cash under memorial circumstances~\cite{kumar2011times,ferreira2014building,vines2012cheque}.
Firstly, the tangibility of cash can serve as a visual aid in some routine workflow~\cite{o2017working}, such as bus ticket buying, which has been chosen and inherited by relevant staff until now. Moreover, such tangibility is seen to improve both payer's and the payee's perceived security during transactions~\cite{pal2018digital}.
Secondly, cash can be credited with additional social implications, which are widely appreciated by society. For instance, Ferria et al. ~\cite{ferreira2014building,Ferreira2017Spending,Ferreira2019SocialConsideration} have elaborated the role of cash in unifying and harmonizing social clusters by rendering necessary interactions among communities. 

Beyond previous reasons for cash preference, the privacy concern amplifies cash use preference. Karin~\cite{Karin2021Michigan} has concluded that cash payment is still favored by people seeking anonymity against digital payment~\cite{Maurer2015Technology}, though such anonymity may bring issues to the government, that the misuse of cash may shelter money laundering, tax evasion, and other illegal activities~\cite{Karin2021Michigan}.


\subsection{Digital-preferred Reasons}

Digital payments are preferred over cash in some scenarios (e.g. pandemic outbreak) or serving as special purposes (e.g. managing investment).


Similarly to the cash's role in personal financial tracking and controlling, digital payments are preferred by households in sophisticated scenarios~\cite{lewis2019follow}, e.g., debt ledge between friends and loan payment. Microsoft has brought up an initiative to solve loan management in an undeveloped country using digital payments~\cite{Mehra2018PrayanaIF}. This initiative turned out to be a success as digital systems have provided people with accurate and timely payment information, which manual methods could hardly achieve. 
Lewis et al.~\cite{lewis2019follow} have suggested that digital payment is more capable and prevalent in financial investment for people who intend to remotely structure wealth and manage their portfolios. The yield of their investment is fluctuating on a daily basis while digital transaction methods could support remote and agile transactions. Investment is currently linked to people's bank accounts as an extra service, which people could access to financial markets, buy in or sell out directly from their savings.

\subsection{Digital-only Reasons}

Nevertheless, digital payments also dominate in scenarios where cash could hardly be accepted. For example, online shopping may require users to finish payment before shipment. Therefore, only digital payment could perform timely transactions ~\cite{adrian2019digital}. Another interesting reason has been discussed in Shen et al. ~\cite{shen2020can} that when merchants experience the advantages of digital payments, they may refuse to accept cash with their customers. As their headline highlights, people are unable to use cash in the most common life scenarios.

Moreover, with the development of the ecosystem of digital payments, digital payment services providers, such as Alibaba and PayPal, provide benefits and conveniences for people who purchase through digital payments~\cite{Guo2016ecosystem}, which are completely inaccessible to people who use cash. Therefore, these incompatible digital payment ecosystems drive people to use digital money when interacting with their services.

\section{Design Space of the Payment System}

As previous studies suggest, people tend to switch between cash and digital payments based on personal preferences or situational factors. While some may prefer cash or digital payments in certain scenarios, other situations may not allow for easy substitution of payment methods. Therefore, we can infer that cash and digital money are integral parts of the current monetary system.

In this section, we propose a payment system that allows users to choose their preferred payment method while also addressing the challenges of offline scenarios where a payment method may not be valid. To achieve this, we follow the framework of the moneywork lifecycle relationships~\cite{perry2018moneywork} and explore the design space of our payment system, focusing on three phases of the money exchange process: pre-transaction, at-transaction, and post-transaction.

\subsection{Unifying Cash and Digital Money}

Our first design principle is to cater to the different preferences of users. During a transaction, the originator and recipient will agree on the form of exchange, as specified in the at-transaction phase of the moneywork framework. However, when parties disagree on payment methods, one of them may need to perform an additional money exchange to obtain their preferred form of currency for housekeeping or sharing, as outlined in the post-transaction phase of the framework.

Currently, the conversion between cash and digital money is heavily reliant on banks. This conversion process typically requires individuals to physically visit a bank or ATM, which can be inconvenient or impossible in certain situations. In our payment system, we aim to optimize this conversion process and make it more accessible to users during the pre-transaction and post-transaction phases, regardless of time or location constraints.

\begin{figure*}
    \centering
    \includegraphics[width=0.9\textwidth]{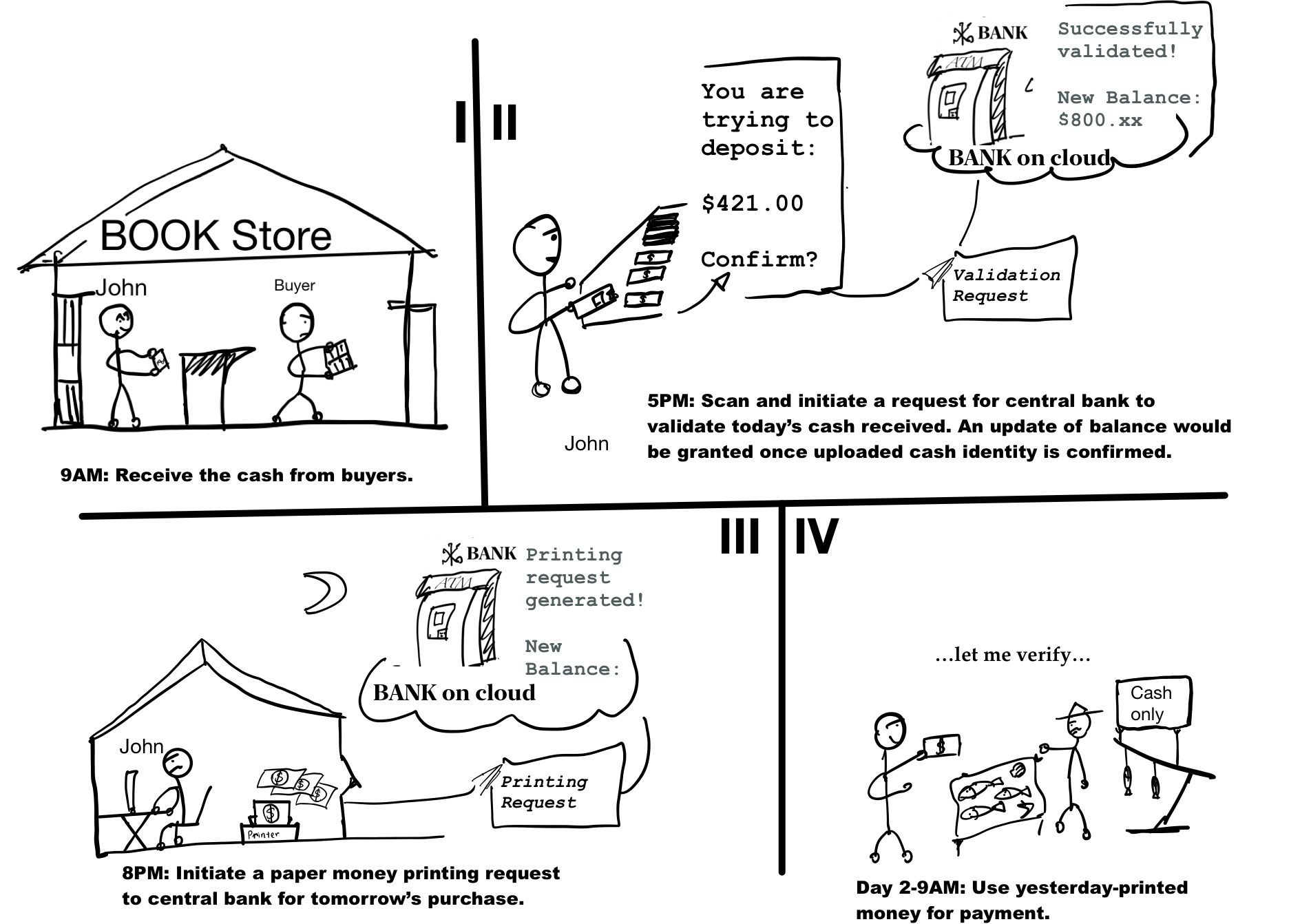}
    \caption{Design sketches - ideal experiences with our payment systems. When people want to deposit the received cash into their accounts, they can upload them to the system by simply taking a picture. Meanwhile, when people need cash, they can print valid cash directly with their personal printers.}
    \label{fig:id1}
\end{figure*}

To achieve this, our system allows for direct acceptance of cash digitally and easy printing of cash from digital accounts. For example, if a user wants to use cash in a digital system, they can deposit it by taking a picture with their mobile phone. Alternatively, if a user wants to obtain cash, they can simply print it themselves instead of visiting a bank or ATM. We illustrate these ideas in Figure \ref{fig:id1}.

To realize this design idea, our system needs to accomplish two requirements. First, the mobile phones (digital devices) should collect the information of cash and validate the authentication of the cash. Second, after the payee confirms the transaction, the original cash should be invalidated. Each piece of cash has its own unique identity and value, and this information is recorded in the system when the cash is generated. The authentication of the cash can be validated in the system, and the information of the cash will be deleted after validation.

\subsection{Exclusive Payment of Cash}

Because our payment system is based on digital payment systems, users can directly use the digital money to conduct transactions when digital payment is the exclusive payment method accepted. The previous design principle aims to address the incompatibility between cash and digital money in the current money system. The system still relies on online validation - although payers can use the printed cash to conduct the payment offline, payees still need to communicate with the online server to validate the cash.

However, the reception of the transaction may not always have a connection to the network during the money exchange process. As shown in the moneywork~\cite{perry2018moneywork}, the last step during at-transaction phase is the confirmation and closure of the transaction, which requires the reception of the transaction to validate the validity of the money. Traditional cash is standardized in terms of its size, weight, and design, made of rigid and light-density material such as polymer or paper. People distinguish the authenticity of cash by their own experience of its physical properties. However, in a digital-based money system, the reception of the transaction should be able to recognize valid money without any physical features. Therefore, we consider addressing this challenge with pre-authorization systems.

\begin{figure*}
    \centering
    \includegraphics[width=0.9\textwidth]{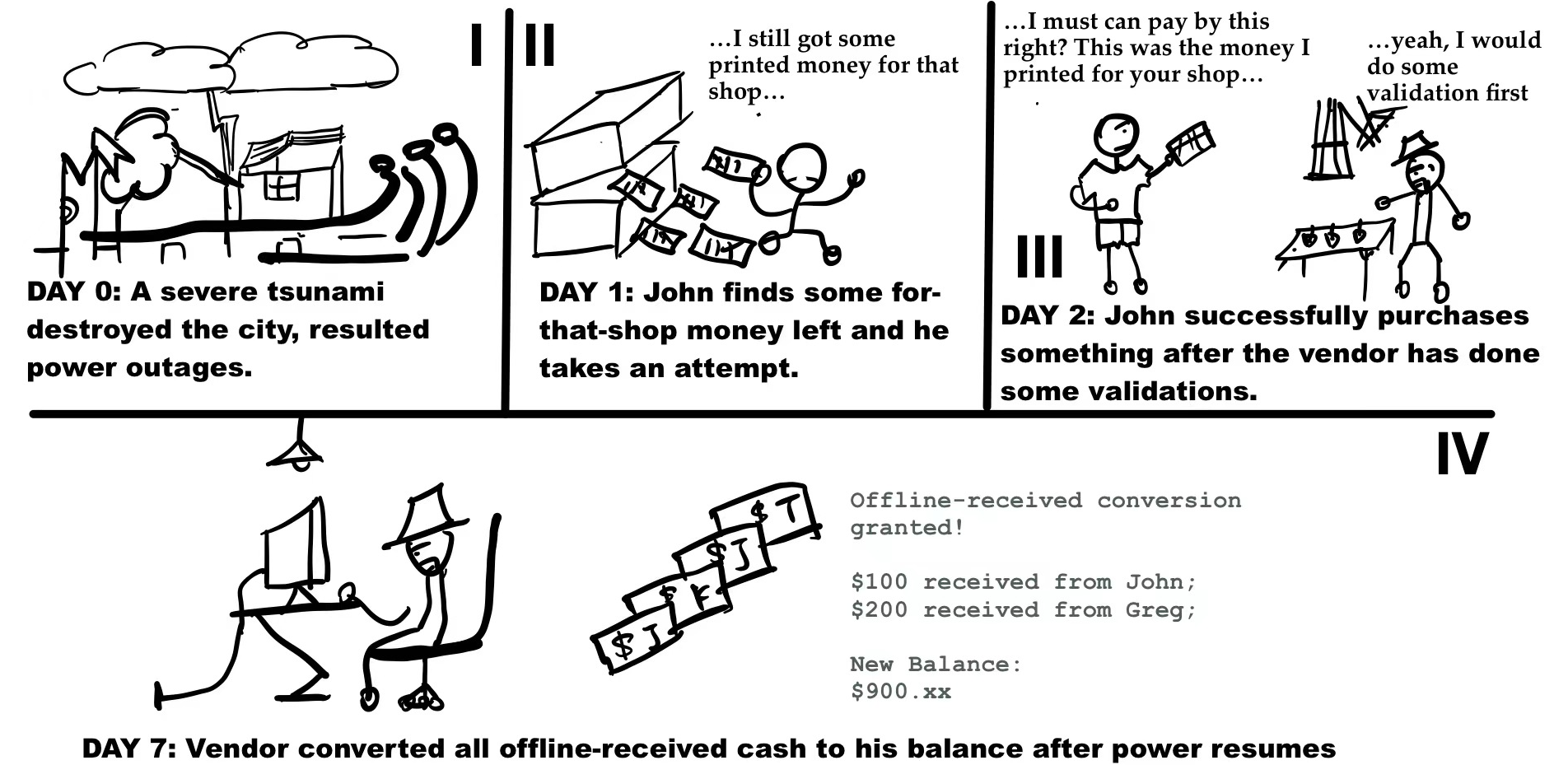}
    \caption{Design sketches - ideal experiences of offline payments. With self-print cash, merchants can verify the validity of the cash without connections to the network.}
    \label{fig:off-post}
\end{figure*}

In current business scenarios, offline verification for merchants only occurs in extreme scenarios, especially in countries and cities where infrastructure is well-developed. Therefore, we do not consider that the merchant pre-authorization systems need to enable offline payment permanently. In other words, users still need to connect to the digital system to update the money status after offline verification. Figure~\ref{fig:off-post} shows an ideal experience of offline payments in an emergency. Users will determine the receiver of the cash before making a payment. The server will generate merchant-specific cash in advance, and the cash can only be used at the specified merchant. Meanwhile, the merchant can verify the validity of the printed cash offline. Afterwards, when the merchants have access to the network, they can update the amount of money that they receive in the system.

To realize this design idea, our system needs to accomplish two requirements in addition to the basic printing money function. First, people cannot manufacture fake money by themselves. Second, people cannot double-spend money at different merchants. Each piece of cash has been encrypted by a unique key of a merchant when it is generated. The cash can only be used with the designated merchants. Because merchants have the records of all offline payments, users cannot spend the same money with the merchants for a second time.

\subsection{Experimental Exploration}

To evaluate our design, we have developed a prototype system, which includes an Android app for users and a centralized server. The details of our prototype are provided in the supplementary file, including the interface and functionality. As we suggested in our design space, to ensure the generality of our money system, we do not require a specific form, material, or presentation of the cash. In our prototype, we provide two cash forms, i.e., QR code and NFT tags, as examples to demonstrate the usability of the money system. For the QR code, users can either print it with a printer or keep it as a photo on their mobile device. For the NFT tags, users need to read/write NFT tags for recording the cash information in their self-generated cash. Apart from the form of the cash, there is no difference between these two variations of cash in our system. Note that although we have not implemented all possible forms of cash in our prototype, other forms of cash are also valid for the design with further development of the system.

At this stage, we test our system in the laboratory, focusing on its usability: Can this system bring convenience to users? Can this system solve the problems that users encounter in real life? What do users think of this new cash system? How do they feel when testing it?

We invite six users to participate in our exploration study. Users first read the system's instructions, try out the system, and ask questions to the researchers. Then, we ask users to complete four tests that cover all the functionality of our system. We record the testing process and collect testing issues. Finally, we discuss their feelings about the payment system with them.

We find that first-time users show a good understanding of the new system. We find two main challenges through our observation of the testing process: network connection and NFC practice. The most significant concern about our system from testing users is security. Several users are curious about recovering the money after a loss or theft and how to track thieves. Our payment system has a remedy plan in case cash is stolen. The system allows users to invalidate the cash they print if it has not been used. The money will then go back to the users' accounts, ensuring that thieves cannot use the cash anymore.

Another concern mentioned by users is managing change money. In our qualitative study, participants also mentioned that in retail scenarios, the cash value does not always perfectly match the amount that payers have to pay to the merchants. Based on different user preferences, we consider two possible mechanisms. The most intuitive solution is to generate new cash by the merchants. The newly printed cash is considered the credential for the change. However, such a mechanism has security pitfalls. Both the merchant and the consumer have the identity and private key of the cash. Therefore, there is a possibility that the merchant embezzles the money. We discuss this security issue with users, and they did not consider it a big problem. Because the system has a record of the cash generation and deposit, they do not worry too much about the merchant stealing the money. Another mechanism is designed to solve the problem proposed by our interviewees who do not want to manage small change. Rather than generating new cash, the change money will directly return to the user's digital account. If the total value of paid cash is equal to or larger than the total amount of the bill, then the system will deposit the bill's amount in the merchant's account and return the change to the account that created the cash. This solution addresses not only the inconvenience of carrying large amounts of small change but also the security issue of the first mechanism.

\section{Discussion}
In this paper, we present the system design of a Central Bank Digital Currency (CBDC). This study aims to provide a better understanding of the design of payment systems, and in this section, we further elaborate on the potential challenges of promoting a payment system and how our system addresses these challenges.

\subsection{Rethinking Current Payment Systems}

Research and surveys indicate that while cash is still widely used and appreciated by people, its usage is decreasing due to the rise of digital payment methods. Cash is a reliable and universally accepted form of payment, particularly in unexpected situations such as power outages. Moreover, certain individuals prefer cash in specific scenarios because it carries face or par value. However, digital payment systems have become faster and more secure in recent years. Despite this, they still lag behind in terms of reliability and sociability when compared to cash. Cash and digital payment methods operate parallel to each other and are only converted by banks after verification. However, this process requires physical deposits at a bank or ATM, which can be challenging for unbanked individuals who may not have access to these services.

We believe that preserving the physical entity of currency is essential for various situations, and the verification process should be somewhat independent of the banking system to facilitate the exchange between digital and cash payments. As a result, our design focuses on preserving the physical attributes of cash while also incorporating the advantages of digital payments. Moreover, by embedding verification information within printable cash, we aim to reduce the need for users to visit banks or ATMs for deposits. This enables individuals to convert cash to digital credit independently and vice versa. Although the physical form of money is still present in the system, it is not entirely anonymous, which preserves authority control over financial activities.

Our system allows for free conversion between digital money and cash, which helps increase financial inclusion by bringing unbanked individuals into the system. This is crucial because financial services should be as accessible as basic necessities like electricity and water~\cite{John2012cheques}. Our system provides various payment forms to cater to different preferences, allowing people to choose between cash or digital payment. As research by Ferria and her colleagues has shown, cash has a strong social aspect as it promotes interaction during transactions~\cite{Ferreira2017Spending,fernandez2020central, ferreira2014building, Ferreira2019SocialConsideration}. Similarly, our system, which incorporates the physical form of cash, will also encourage positive interaction during offline transactions. Additionally, the seamless conversion between cash and digital payment eliminates switching costs between two payment systems, leading to a more overall hybrid payment pattern where people can continue to use their preferred method while being encouraged to try new methods.

Notwithstanding, we have also noticed two challenges that exist for users to adapt to a new digital payment system: 1) implementation challenges that may resist the system's promotion; 2) security challenges regarding how people could adopt and trust our system. We will address these two aspects accordingly in the rest of this section.

\subsection{Implementation Challenges for Promoting Payment Systems}

There are three implementation challenges that current system design faces: 1) costs from the process of printing one's money; 2) costs of other devices and infrastructure; 3) increasing pressure imposed on the bank.

Our exploration has demonstrated that using self-printed cash to transmit payment information is valid. However, this presents a challenge as it generates a significant amount of paper waste and can have a negative impact on the environment. Additionally, the returns of using cash may be diminished if the cost of printing it is too high. However, our system is more cost-effective compared to traditional cash systems~\cite{usdollarcost}. Unlike traditional cash, which needs to be verified through physical means and can circulate for long periods, our self-printed money does not require special materials and can be authenticated through the information it carries~\cite{usdollarpr}. The use of NFC chips or other trusted mediums is acceptable for this purpose. Furthermore, with the advent of digital payment systems, people may prefer to deposit the self-printed money they receive into their accounts, thus shortening the circulation period and reducing the cost of printing money. While it may seem like the cost is shifted to end-users, governments can compensate individuals who use the system.

Another challenge in our prototype system is the accessibility of self-printed money. The use of external devices, such as printers, may limit the flexibility and versatility of the system. However, it is important to note that the portable printer in our prototype system is just one example and not the only solution for self-printed cash. While the system does require users to have access to paper and printers, it may still be more accessible than traditional ATM machines and bank branches, which are becoming increasingly scarce~\cite{Swe}. We do not claim that our system is superior to the current bank system in all cases. In fact, we envision the current bank system to be integrated into our money system as a public resource for those who do not have their devices or are in emergencies. For example, people can still print their money at bank branches and ATM machines as before. Additionally, as self-printed cash is cheaper than traditional cash, central banks and governments may assist in setting up the system's infrastructure to ensure better accessibility.

Central systems may experience a significant surging demand for authentication from users, such as requests for printing, authentication for money received, and balance retrieval. This exposes the bank's network to great vulnerability, making it difficult to integrate cash payments into the digital payment system while keeping it available. For example, if the service is down, it may be challenging to verify printed cash. Overall, designing the system architecture for banks will be challenging and transformative. Currently, our work focuses on the interaction design between users and payment systems, but we have not yet studied the back-end system architecture or tested its performance. Other researchers have explored different implementations of CBDC systems, such as using distributed ledgers, and other financial sectors may also be involved in the CBDC system to help share the workload of the central system~\cite{lbcoin, yao2018, yao20182}.

\subsection{Security Challenges in Endorsing Payment Systems}

One major challenge in our payment system is ensuring that end-users trust self-printed money. While cash can be lawfully accepted and liquidated for payment if it is real, payees must be able to verify payment both online and offline in order for a payment system to be widely used. Therefore, we will explore the various trust spectrums of payment systems at different levels and address payee concerns.

At the lowest security level, a QR code represents money in the payer's account, and whoever scans the QR code first will receive the money. This is similar to a personal check in the physical world. Personal checks may be acceptable for close relationships where there is already a high level of trust, but may not be accepted by retail establishments. In an emergency scenario, a personal check could be carried in an NFC-enabled software like Google Pay or in printed form.

At the next security level, the paying account belongs to a trusted third party, such as a commercial bank, and the equivalent in the physical world is a banker's check. A retail store may accept such a check even if the customer is unknown, as it is guaranteed to be valid by the bank that issued the check. While a fraudulent customer may attempt to double-spend by copying the check, the issuing bank will pursue such fraudulent activity. Furthermore, a banker's check can be made more secure with printed security elements, similar to banknotes. Banks may be willing to issue digital money in the form of such banker's checks as a business.

At the highest security level, colored money in the cryptocurrency world is akin to a gift card of a particular store in the physical world. The money is printed so that there is only one possible receiver, such as a retail store, which ensures that there is no double-spend. In this sense, the money was already promised to the store when it was printed, and the customer only waits for an opportunity to spend it. Unlike traditional gift cards, colored money can be reverted back into the payer's account.


\section{Conclusion}

This paper addresses the challenges arising from the development of digital payments, especially with regard to central bank digital currency (CBDC) systems. The question we aim to answer is: how can we design CBDC systems that enable all members of society to access financial services as easily as they access public utilities? To answer this question, we propose a CBDC system with self-printed cash that is compatible with digital payment systems and resilient to emergency scenarios. Our solution presents a new approach to the development of CBDC systems and could serve as a blueprint for the design of hybrid digital currencies.

\begin{acks}

This work was supported in part by the University of Macau grants SRG2022-00032-FST	
 and MYRG-CRG2022-00013-IOTSC.
\end{acks}

\bibliographystyle{ACM-Reference-Format}
\bibliography{sample-base}

\end{document}